# An FFT-solver used for virtual Dynamic Mechanical Analysis experiments: Application on a glassy/amorphous system and on a particulate composite.

Stéphane André[(1,2)], Julien Boisse[(1)], Camille Noûs[(2)]

(1) Université de Lorraine, CNRS, LEMTA, F-54000 Nancy, France
(2) Cogitamus Laboratory, F-75000, France

**ABSTRACT:**

FFT-based solvers are increasingly used by many researcher groups interested in modelling the mechanical behavior associated to a heterogeneous microstructure. A development is reported here that concerns the viscoelastic behavior of composite structures generally studied experimentally through Dynamic Mechanical Analysis (DMA). A parallelized computation code developed under complex-valued quantities provides virtual DMA experiments directly in the frequency domain on a heterogenous system described by a voxel grid of mechanical properties. The achieved precision and computation times are very good. An effort has been made to show the application of such virtual DMA tool starting from two examples found in the literature: the modelling of glassy/amorphous systems at a small scale and the modelling of experimental data obtained in temperature sweeping mode by DMA on a particulate composite made of glass beads and a polystyrene matrix, at a larger scale. Both examples show how virtual DMA can contribute to question, analyze, understand relaxation phenomena either on the theoretical or experimental point of view.

# 1. Introduction

Dynamic Mechanical Analysis (DMA) is known to be a privileged tool to study materials - especially polymers and rubbers- whose rheological behavior is viscoelastic by nature, i.e. introduces irreversible dissipation of mechanical energy into heat. The technique indeed measures a conservative (storage) or dissipative (loss) modulus or compliance, which are the real and imaginary parts of their complex nature $M^*(\omega) = M'(\omega) + iM''(\omega)$ for instance for the modulus. It relies on harmonic steady-state excitations in strain, applied at pulsation $\omega$ on a material specimen and on the recording of the corresponding output stress signal. This latter is analyzed with respect to the original input both in amplitude decrease (modulus $|M^*|$) and phase lag $\delta$ (reflecting the damping phenomenon associated to some viscous component of the behavior). From that information, complex algebra gives eventually access to $M'(\omega), M''(\omega)$ with a sweep in frequency allowing for full dynamical characterization of the material. Such a spectroscopic probing of the material exists in many different fields of physics like in thermal science to produce thermal conductivity/diffusivity measurements (Cahill 1990) but the main corpus of publications probably resorts to dielectric properties measurements (Dissado 2017). Experimental data on the susceptibility of physical processes as function of frequency is, in a general way, a central tool to develop physical model of relaxation processes (Dissado and Hill 1987, Havriliak and Havriliak 1995, Jonscher 1996)

DMA is estimated 100 times more sensitive to glass transition than scanning calorimetry (Menard and Mennar 2015) and this is probably the reason why the sweeping temperature mode is generally elected by material scientists as the perfect way of investigating subtle microstructural transitions (motions of polymer chains for example) (Wiley 2008). But for solid-liquid rheological characterization, the forced frequency sweep is more desirable as it allows behavior's law modeling assumptions to be checked and associated material parameters to be identified. The targeted information is obviously the relaxation spectrum which can be precisely investigated with the frequency scanning. Emri and Tschoegl (Emri and Tschoegl 1995) synthetize the spectrum from temporal data in a collocation-like approach but check the consistency with DMA results in frequency mode. In Kim and Lee (Kim and Lee 2009) on the contrary, experimental characterization of Frequency Response Functions (FRF) are used to identify parameters originating from constitutive rheological models of damping materials. One key issue of this study precisely results in the solution offered by virtual DMA to provide quickly these data for any kind of material and to test theoretical concepts underlying such models.

It should be pointed that this technique is before anything else a perfect tool to investigate the behavior of a material at the scale of a Representative Element of Volume (R.E.V.). Samples are generally of small (macroscopic) sizes and excitations of low magnitude (small perturbations theoretical framework). Independently of excitation modes and specimen geometry (torsion, flexure, compression), this allows retrieving local material parameters directly from rheological models. These parameters are the exact reflect of statistically averaged microstructural evolutions.

The concern of this paper is classical in scientific calculus: offering a simulation path to replace experiments and exceed their intrinsic limitations or, in other words, to offer a virtual DMA simulator capable of analyzing any kind of heterogeneous or composite material. Such simulation tool has been already striven after in the past through Finite element approach. Brinson and Knauss (Brinson and Knauss 1992) for example have modified a FE code to make it possible the computation of real and imaginary complex moduli directly, solving the boundary value problem with complex variables.

New computational tools, more efficient, have made recently possible to rejuvenate this idea thanks to two important advances in computational science:

i. synthesis of virtual composite microstructure using generators working on various mathematical basis (Salnikov et al. 2015, Ghazvinian et al. 2014, Quey 2014) or alternatively synthesis of real microstructure based on tomography imaging followed by appropriate image treatment (Uchic et al. 2007). Virtual DMA performed with FE codes was initially limited to very simple idealized microstructures.

ii. the development of the spectral approach (Fast-Fourier Transform operator) to handle the resolution of local equilibrium equations (Moulinec and Suquet 1998, Roters et al. 2019 for a wide review on the topic, especially for works developed using the Damask code). Of course, the harmonic steady-state regime can be obtained from time domain simulations based for example on Finite Element approach (Masurel et al. 2015), but such approaches are very much less efficient than the spectral ones directly operating in frequency domain.

The idea is before anything else to make the confrontation of both approaches (simulation and experiment) a source of better knowledge arising from their respective drawbacks.

- Simulation will always be limited by the idealization of the composite organization: for example, if one can faithfully and easily reproduce particulate composite by considering real size distributions of particulate, real volume fractions…, continuity assumptions are generally considered between particles and the matrix. Overcoming this complexity is gaining attention however. An homogenization problem in thermal science has recently addressed the case of imperfect interfaces within the FFT spectral approach (Monchiet, 2018).
- Experiments based on DMA can carry a bias inherent to the technique: as an example, one can cite the drift in measurement signals with the very long-time durations of these experiments to get low frequency information.

Of course, the second key idea is the homogenization problem of composite materials. It is in that direction that very recent computations of virtual DMA kind have been performed with FFT-solvers. In a probably landmark article of 2016, Figliuzzi (Figliuzzi et al. 2016) studied the composite material made of a rubber matrix, filled with carbon black fillers. The effective mechanical behavior resulting from various models considered for the multiscale morphological microstructure description has been successfully computed in 3D with an FFT-solver in space and harmonic complex treatment of the dynamic part. The frequency dependence of the effective complex bulk modulus and of the effective shear modulus was directly achieved and shown as a valuable way to assess the performance of analytical effective models. A second study of that type was used by Gallican (Gallican and Brenner 2019), to investigate the overall properties of composite materials with fractional viscoelastic constituents. In the case of particulate composites with polydisperse spherical elastic inclusions, the authors develop FFT based computations of the overall complex moduli in frequency domain. Such outputs were helpful to assess mean-field approximate models and a generalized fractional effective model derived from exact mathematical (asymptotic) relations constraining the adjustment of an effective relaxation spectrum.

It is worthwhile to mention that whatever the virtual DMA computation tool (FEM or FFT-based solvers), a comfortable aspect is that viscoelastic theories in the linear regime require only infinitesimal deformation analysis for strains and displacements.

The results presented here enter into this movement and favors an account of the benefits that can be expected from virtual DMA to explain physical mechanisms or question theoretical model assumptions (section 3.1) as well as experimental data (section 3.2). In section 2 will be firstly described the extension to the harmonic steady-state regime of spectral FFT solvers used to solve local equilibrium equations generally in the steady state or for temporal responses. Validation of the numerical implementation will consider the academic heterogeneous checkerboard structure with constituents of Standard Linear Solid (SLS) rheological behavior. Additionally, it will illustrate how outputs of virtual DMA can be used to identify material transfer functions, a concept which will allow to compute the response of the material to any kind of solicitation. Section 3 will be devoted to illustrate what a virtual DMA solver can bring to material engineering science in the future. Two test cases will be considered, each of them showing a different aspect of the subject. Example 1 will compare the results based on the paper of Masurel et al. (Masurel et al. 2015) for a multi-material made of a collection of Maxwell units and obtained by time-domain calculations using the FE code Zebulon. Finally, example 2 will illustrate the benefit of using virtual DMA when trying to interpret experimental results. It will rely on data relative to strong experimental works made by Alberola and collaborators (Agbossou et al. 1993, Alberola and Mele 1996) on a Particulate Composite of glass beads in a polystyrene matrix. It shows also that harmonic effective behavior can be calculated directly from the knowledge of either the temperature or frequency-discretized moduli of the constituents. This "option" can be very useful for experimentalist where DMA data are often produced at a given frequency with a sweep in temperature.

## 2. An FFT-Spectral solver extension to harmonic regime

2.1. Standard mathematical procedure and numerical implementation

We consider a heterogeneous multi-material system, with a Representative Volume Element (RVE) $V$, subject to oscillatory boundary conditions with angular frequency $\omega$. The heterogeneous harmonic displacement field $\tilde{\mathbf{u}}(\mathbf{r}, t)$ in such harmonic regime can be expressed with a complex form:

$$\tilde{\mathbf{u}}(\mathbf{r}, t) = \mathbf{u}^*(\mathbf{r}) e^{i\omega t} \tag{1}$$

$\mathbf{u}^*(\mathbf{r})$ with the $*$ upperscript denotes the complex displacement amplitude, whose real part corresponds to the physical displacement field. Considering the harmonic regime, and in the small perturbation approximation, the spectral method based on the Lippman-Schwinger equation associated to a discrete Green operator is straightforward. The complex amplitudes of the strain $\boldsymbol{\varepsilon}^*(\mathbf{r})$ and stress $\boldsymbol{\sigma}^*(\mathbf{r})$ are second rank tensors, which depend on the frequency $\omega$.

Now considering a heterogeneous medium made of linear viscoelastic phases and thanks to the correspondence principle (Hashin 1970), local constitutive laws expressed below in terms of relaxation functions can be put directly in the complex form (Laplace-Carson transforms in time where the Laplace variable $p$ is set to $i\omega$), with a linear relation:

$$\boldsymbol{\sigma}^*(\mathbf{r}, i\omega) = \mathbf{H}^*(\mathbf{r}, i\omega) : \boldsymbol{\varepsilon}^*(\mathbf{r}, i\omega) \tag{2}$$

The complex local stiffness tensor $\mathbf{H}^*(\mathbf{r}, i\omega)$ accounts for the storage and loss moduli.

In the frequency (spectral) domain, the local equations expressing the constitutive law (Eq. 2), the mechanical equilibrium, and the overall prescribed strain loading produce the system:

$$\begin{cases} \boldsymbol{\sigma}^*(\mathbf{r}, i\omega) = \mathbf{H}^*(\mathbf{r}, i\omega) : \boldsymbol{\varepsilon}^*(\mathbf{r}, i\omega) & \forall \mathbf{r} \in V \\ \mathbf{div}\, \boldsymbol{\sigma}^*(\mathbf{r}, i\omega) = \mathbf{0} & \omega \in [0; +\infty[ \\ \langle \boldsymbol{\varepsilon}^*(i\omega) \rangle = \bar{\boldsymbol{\varepsilon}}^* \end{cases} \tag{3}$$

It can be solved by making use of a numerical scheme based on the Fourier Transform (Figliuzzi et al. 2016, Moulinec and Suquet 1998) considering the RVE with periodic boundary conditions. The elastic classical scheme introduced by Moulinec (Moulinec and Suquet 1998) has to be transposed in the Laplace-Carson domain with complex-valued quantities. The Lippman-Schwinger equation is transformed in the Fourier domain ( $\widehat{\phantom{x}}$ symbol) and depends now on spatial frequencies $\mathbf{k}$ and dynamical testing frequencies $\omega$:

$$\hat{\boldsymbol{\varepsilon}}^*(\mathbf{k}, i\omega) = -\hat{\boldsymbol{\Gamma}}_0^*(\mathbf{k}) : \hat{\boldsymbol{\tau}}^*(\mathbf{k}, i\omega) \qquad \forall \mathbf{k} \neq \mathbf{0}, \quad \hat{\boldsymbol{\varepsilon}}^*(\mathbf{0}, i\omega) = \bar{\boldsymbol{\varepsilon}}^*, \quad \omega \in [0; +\infty[ \tag{4}$$

$\hat{\mathbf{\Gamma}}_0^*$ is the periodic Green's operator in Fourier domain, associated to the reference medium of isotropic stiffness tensor $\mathbf{H}_0^*$, which has to be real (Figliuzzi et al. 2016). $\hat{\boldsymbol{\tau}}^*(\mathbf{k}, i\omega)$ is the Fourier transform of the polarization field $\boldsymbol{\tau}^*(\mathbf{r}, i\omega) = \boldsymbol{\sigma}^*(\mathbf{r}, i\omega) - \mathbf{H}_0^* : \boldsymbol{\varepsilon}^*(\mathbf{r}, i\omega)$. $\bar{\boldsymbol{\varepsilon}}^*$ denotes the average over the whole R.E.V. (Fourier frequency $\mathbf{k}=0$) of the local complex strain field $\boldsymbol{\varepsilon}^*(\mathbf{r}, i\omega)$. Equation (4) can be solved in Fourier domain (spatial discretization) and for the harmonic regime described by any angular-frequency set $\in [0; +\infty[$ and with fully complex local $\boldsymbol{\varepsilon}^*(\mathbf{r}, i\omega)$ and $\boldsymbol{\sigma}^*(\mathbf{r}, i\omega)$ fields. A numerical scheme is necessary, of fixed-point type, to obtain convergence along an iterative process over the polarization field variable, where iterate $n + 1$ is updated from the previous iterate $n$. High-performance accelerated numerical schemes are available for this purpose (Eyre and Milton 1999, Moulinec and Silva 2014)

A convergence test is computed at each iteration. It consists in checking the deviation from equilibrium $e_{equ}$, from compatibility $e_{comp}$ and from the prescribed loading conditions $e_{load}$. The iterative procedure is stopped when these 3 criteria are smaller than some prescribed values (most of the times fixed to $10^{-4}$).

The deviation from equilibrium is obtained with the following dimensionless criterium, calculated in the Fourier space:

$$e_{eq(uilibrium)} = \frac{\|div(\boldsymbol{\sigma}^*) \, dV\|}{\|\boldsymbol{\sigma}^* . \boldsymbol{n} \, dS_n\|} = \frac{\sqrt{\sum_k \hat{\mathbf{f}}_{eq} \cdot \hat{\mathbf{f}}'_{eq}}}{\sqrt{\sum_k \{\hat{\mathbf{f}}_1 \cdot \hat{\mathbf{f}}'_1 + \hat{\mathbf{f}}_2 \cdot \hat{\mathbf{f}}'_2 + \hat{\mathbf{f}}_3 \cdot \hat{\mathbf{f}}'_3\}}} \quad (5)$$

with the following expression of the forces (the prime symbol refers to a complex conjugate).

$$\hat{\mathbf{f}}_{eq}(\mathbf{k}) = \hat{\boldsymbol{\sigma}}^*(\mathbf{k}) : i\mathbf{k} \, . \, dV$$
$$\hat{\mathbf{f}}_1(\mathbf{k}) = \hat{\boldsymbol{\sigma}}^*(\mathbf{k}) : \mathbf{n}_1 \, . \, dx_2 dx_3$$
$$\hat{\mathbf{f}}_2(\mathbf{k}) = \hat{\boldsymbol{\sigma}}^*(\mathbf{k}) : \mathbf{n}_2 \, . \, dx_1 dx_3$$
$$\hat{\mathbf{f}}_3(\mathbf{k}) = \hat{\boldsymbol{\sigma}}^*(\mathbf{k}) : \mathbf{n}_3 \, . \, dx_2 dx_1$$

The deviation from compatibility is obtained with the following expression calculated in the Fourier space

$$e_{comp} = \frac{\max_{\mathbf{k}} \left( \max_{j=1,\dots,6} (|\hat{e}_j(\mathbf{k})|) \right) . \max(L_1, L_2, L_3)^2}{\sqrt{\sum_{\mathbf{k}} \hat{\varepsilon}^*_{mn}(\mathbf{k}) . \hat{\varepsilon}^{*\prime}_{mn}(\mathbf{k})}} \quad (6)$$

where $L_i = N_i dx_i$ is the size of the box in directions $i = 1, 2, 3$ and with $\hat{e}_j(\mathbf{k})$ expressed for the 6 compatibility equations in terms of the tensorial components $\hat{\varepsilon}^*_{mn}$

In the case of a prescribed macroscopic strain $\bar{\varepsilon}_{ij}^*$, the deviation from the prescribed loading is given by

$$e_{load} = \frac{\|\langle \varepsilon^* \rangle - \bar{\varepsilon}^*\|}{\|\bar{\varepsilon}^*\|} = \frac{\sqrt{\left(\langle \varepsilon_{ij}^* \rangle - \bar{\varepsilon}_{ij}^*\right):\left(\langle \varepsilon_{ij}^* \rangle - \bar{\varepsilon}_{ij}^*\right)'}}{\sqrt{\bar{\varepsilon}_{ij}^* \bar{\varepsilon}_{ij}^{*'}}} \quad (7)$$

Other loading conditions, like prescribed stress or multi-directional loadings, can be treated in a similar way. In the Eyre and Milton scheme, and considering a multi-material system with M linear viscoelastic phases the optimal choice for bulk and shear moduli for the isotropic reference medium is :

$$\begin{aligned}\mu_0 &= \sqrt{\max_{i,\omega}(\mathcal{R}e[\mu_i(i\omega)]).\min_{j,\omega}(\mathcal{R}e[\mu_j(i\omega)])} \quad i \neq j \quad i,j \in 1,\dots,M \\ K_0 &= \sqrt{\max_{i,\omega}(\mathcal{R}e[K_i(i\omega)]).\min_{j,\omega}(\mathcal{R}e[K_j(i\omega)])} \quad \omega > 0 \text{ and } \omega \in [\omega_{min}; \omega_{max}]\end{aligned} \quad (8)$$

This mathematical procedure has been implemented as a new option in a spectral solver software named CraFT for Composite response and Fourier Transform developed at LMA laboratory (http://craft.lma.cnrs-mrs.fr) by several scientists since the initial version raised by Moulinec and Suquet (1998). This option takes advantage of all recent developments of the solver in terms of accelerated convergence schemes and is available in parallelized configuration for enhanced computation times. The "harmonic" version is simply available by selecting an appropriate compiling option which proceeds with a set of functions that have been modified to handle the complex nature of the variables (65 functions over a total of 263 including the behavior's law implementation and solver operations). The main change with respect to the "temporal" version lies in the specification of frequency vectors in the Fourier domain due to the two-sided character of the Discrete Fourier Transform. The code modifications render it even more regular in terms of tuplet management. Indeed, for a real function, let's say a stress tensor component, the 3D *double* array of dimension $n_0 \times n_1 \times n_2$ (indices in the three directions determining which voxel is considered) produces a DFT output of $(n_0/2 + 1) \times n_1 \times n_2$ *complex* numbers. (http://www.fftw.org/fftw3_doc/Real_002ddata-DFT-Array-Format.html#Real_002ddata-DFT-Array-Format). In the harmonic version, we have to handle 3D data arrays of complex numbers which restore a simpler approach of the code. Additionally, the equilibrium criterion in Fourier domain Eq. (5) has been implemented, which is formulated in terms of a true normalized force balance (volumic over surface for each voxel). In comparison with the original CraFT criterion based on the stress tensor divergence presenting a dimensional dependence, this results in a unique criterion specification.

Spectral solvers imposed themselves as an alternative approach for computing mechanical problems on heterogeneous structures with the following recognized advantages: precision and very fast computational times. Note that for stochastic materials, the latter is a crucial one as multiple simulations are required for identical statistical figures to obtain an averaged meaningful response of the heterogenous RVE. Among the drawbacks, one can cite two of them, being already subjected to a common proposal for reducing their nuisance: (i) spectral solvers can produce high frequency oscillations on the mechanical fields close to interfaces; (ii) because the voxel distribution is fixed once (no remeshing), the precision of the method can deteriorate rapidly in the case of complicated interfaces and/or multiple length scales. For both drawbacks, the composite voxel approach offers substantial improvements: a filtering effect for the first drawback (Gélébart and Ouaki 2015) and more accurate predictions in the latter case (Mareau 2017, Charière et al. 2020).

2.2. Numerical Validation: the checkerboard case

We consider the academic case of the checkerboard composite, classically used for validating the FFT simulations (Lebensohn 2005). The microstructure is made of a periodic 2D, 2-phase composite whose unit cell consists of four "tiles" with the "crystallographic" orientations of the two pairs of opposite grains being at +90° and 0°. Each material has a rheological behavior described by the 3-parameters SLS solid

$$\left.\frac{\bar{\sigma}(s)}{\bar{\varepsilon}(s)}\right|_{s=j\omega} = G^{(k)}(j\omega) = \frac{G_r^{(k)} + G_g^{(k)} j\omega\tau^{(k)}}{1 + j\omega\tau^{(k)}} \quad k = 1,2 \qquad (9)$$

Where $G_g$, $G_r$, $\tau$ denotes respectively the glass (instantaneous) modulus, the relaxed modulus and a relaxation time. The effective behavior of this checkerboard structure with 2 constituents and 2 orientations submitted to an overall shear excitation has an analytical solution. The Hashin principle states that

$$\tilde{G} = \sqrt{G^{(1)}(j\omega)\, G^{(2)}(j\omega)} \qquad (10)$$

Hence

$$\tilde{G} = \tilde{G}_g \sqrt{\frac{\left(j\omega + \frac{1}{\tau_z^{(1)}}\right)\left(j\omega + \frac{1}{\tau_z^{(2)}}\right)}{\left(j\omega + \frac{1}{\tau^{(1)}}\right)\left(j\omega + \frac{1}{\tau^{(2)}}\right)}} \qquad (11)$$

with $\tilde{G}_g = \sqrt{G_g^{(1)} G_g^{(2)}}$ and $\frac{1}{\tau_z^{(k)}} = \frac{G_r^{(k)}}{G_g^{(k)} \tau^{(k)}}$  $\tau_z^{(k)}$ figures the relaxation time which appears in the zeros of the rational function in Eq.(11). Note that we will also have $\tilde{G}_r = \sqrt{G_r^{(1)} G_r^{(2)}}$.

Real and imaginary parts of the effective complex modulus $\tilde{G}$ can be calculated exactly from Eq. (11) and directly compared to the numerical FFT-solver outputs in virtual DMA mode. Figure 1 below plots the results in the case of a strong contrast in mechanical properties (same values as those in Suquet, 2012). Table 1 below synthesizes the input and output data.

It is clear from Fig.1 that the virtual DMA computations are in perfect agreement with the exact solution. One of the virtual DMA advantages can already be illustrated here: inverse parameter identification (i.e. metrology of material parameters) can be applied to virtual DMA outputs, providing an effective model has been selected. From theoretical considerations (Gallican et al., 2017), it is known that the effective material built from two SLS-type constituents does not follow this rheological behavior itself. This is clearly evidenced by figure 1. The 2-order of magnitude ratio between $\tau^{(1)}, \tau^{(2)}$ implies a marked spectrum of the dissipative (loss) modulus. An effective model could be proposed through the collocation method applied with a two-terms Prony series. This leads in Laplace domain to a rational admittance $Y(j\omega)$ having 2 poles (the relaxation times) and one zero.

$$Y(j\omega) = G_g + (G_r - G_g)\frac{(1 + j\omega\tau_{z_0})}{(1 + j\omega\tau_{p_1})(1 + j\omega\tau_{p_2})} \quad (12)$$

|  | Instantaneous (glassy) modulus | Relaxed modulus | Relaxation times | | | |
|---|---|---|---|---|---|---|
| Constituent 1 | 10 MPa | 9 MPa | 5 s | | | |
| Constituent 2 | 100 MPa | 50 MPa | 0.05 s | | | |
| | | | | | | |
| Effective Exact $\tilde{G}$ (Eq.23) | $\tilde{G}_g = \sqrt{1000}$ | $\tilde{G}_r = \sqrt{450}$ | $\tau^{(1)} = 5s$ | $\tau^{(2)} = 0.05s$ | $\tau_z^{(1)} = \frac{50}{9} s$ | $\tau_z^{(2)} = \frac{1}{10} s$ |
| Effective identified model (2-term Prony series) | $\tilde{G}_g = 31.62$ | $\tilde{G}_r = 21.21$ | $\tilde{\tau}_{p1} = 4.97s$ | $\tilde{\tau}_{p2} = 0.059s$ | $\tilde{\tau}_{z0} = 4.42s$ | |

Table 1: Input/Output values of the checkerboard test case.

Table 1 gives the virtual measurements of the material parameters of this model as resulting from a least-square minimization of the residuals between a (here virtual) DMA experiment and the candidate effective model (Eq.12). The agreement is perfect which means that the mathematical structure of the collocation model (Biot model) catch all the physical content of the signal and would be able to predict the response of the composite material to any kind of input excitation. The moduli $\tilde{G}_g$ and $\tilde{G}_r$ are perfectly recovered and the relaxation times (the two poles of the admittance fraction in this case) are shown to be very close to those of the constitutive materials.

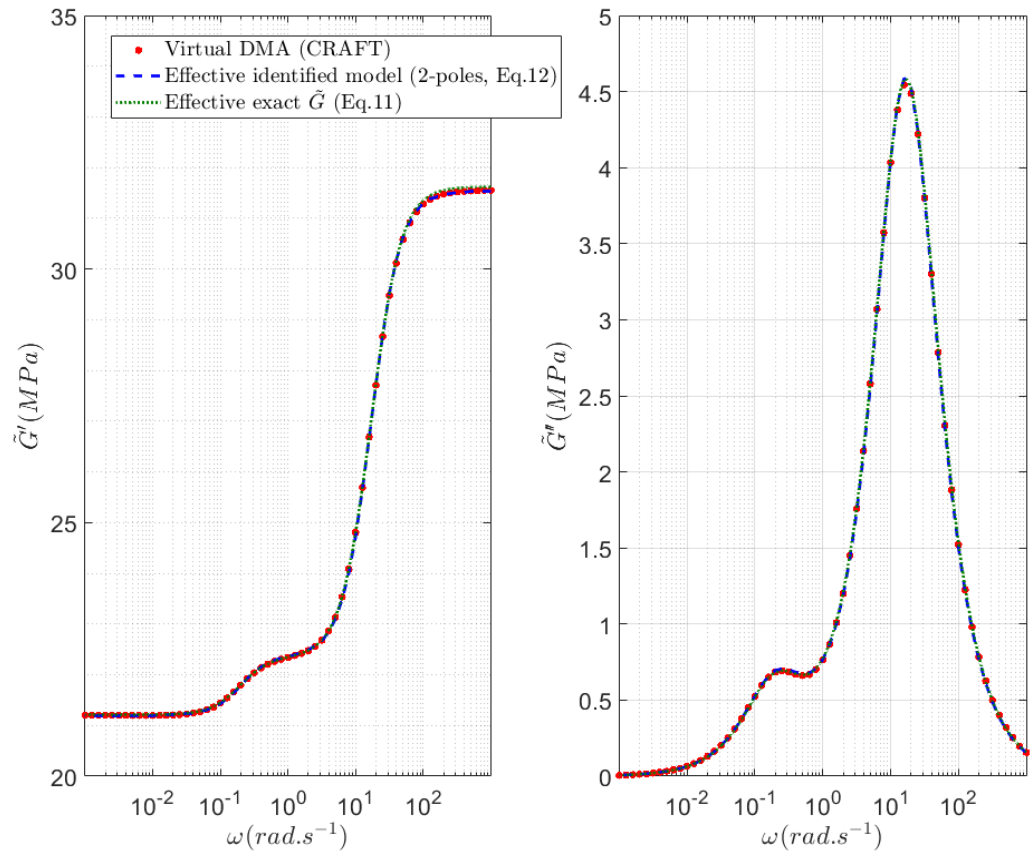

Figure 1: Validation of CRAFT virtual DMA implementation by reference of the exact effective real (storage) and imaginary (loss) moduli as function of frequency.

## 3. Virtual DMA Applications and Interest

3.1: Application 1: Heterogeneous dynamics in Glass/amorphous Polymers

### 3.1.1 The heterogeneous problem

The composite material under concern in this section was studied by Masurel and coll., (Masurel et al. 2015) as an academic approach to understand dynamical heterogeneities which take place in amorphous polymers in the glass transition. We named it a multi-material composite as it is built from a grid of elements with individual behavior being of Standard 3-parameter Maxwell rheological type (Fig.2 -Left). It can be considered as a kind of construction of an "averaged" object representing the entire system at all sub-scales. Observed relaxation processes in disordered systems reveal a kind of universal nature, with now well identified features and/or explicative ideas (cooperative dynamics, cluster formations, energy–criterion arguments…). This multi-material composite aims at catching some of these general features from basic ingredients that are used in a composite view. This is fully motivated by experimental evidences that non-exponential responses of a macroscopic sample results for example from a relaxation-rate distribution of independently randomly sized mesoscopic regions (Jurlewicz and Weron 2002).

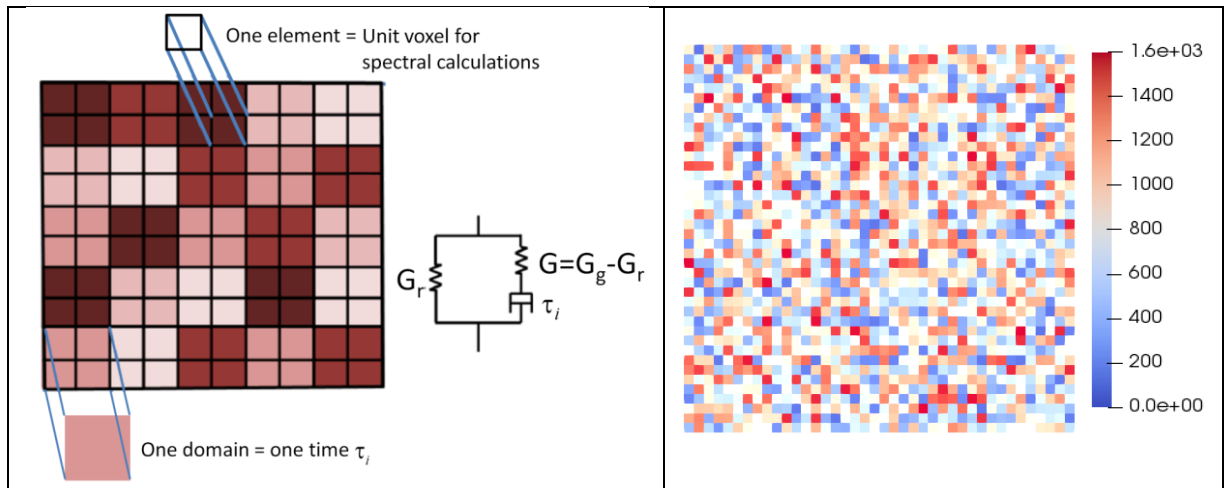

Figure 2: (Left) Domains of the heterogeneous multi-material having a Standard Linear solid Maxwell rheological behavior with all glassy and rubber moduli identical but different, randomly distributed, relaxation times $\tau_i$. (Right) Example of a sampled material repartition of relaxation times in $40 X 40 = 1600$ voxels for virtual DMA simulation.

Here the domains are of identical sizes but dynamics on each domain are randomly distributed. A few identical elements can be associated to form a domain $i$ having a constant relaxation time $\tau_i$ (same mechanical response) and domains are associated randomly to build the polymer *coarse-grained* model. The spectrum of relaxation times has been chosen according to a normal distribution of width $s$ with maximum centered on $\tau^{max} = 1s$ and in logarithmic scale. The glassy and rubber elastic constants ($G_g$ and $G_r$ respectively), being identical for all domain, they applied then to the homogenized composite. This is a strong assumption of the approach which has the advantage to isolate the only contribution of heterodynamics to the homogenized relaxation spectrum (if any).

The objective of the study was to observe the effective behavior of such model in terms of effective relaxation spectrum as well as the local mechanical fields interplay. To this aim, FE simulations were performed using Zebulon code [Z-set: http://www.zset-software.com/] in 2D plane strain condition (frame axis 1,2), assuming incompressibility with a bulk modulus fixed to $K = 10^5 G_g$, periodic boundary conditions, and a pure shearing test (imposed macroscopic shear step strain $\varepsilon_{12} = 0.01$).

We must point here that the study produces <u>time domain</u> simulation and <u>therefore does not give a direct way of computing the complex, frequency dependent, modulus $G * (\omega)$</u>. The authors (see details in Masurel et al. 2015) are led to a post-processing the simulated data in order to identify the frequency complex modulus from the time response of the macroscopic $\langle \sigma_{12}(t) \rangle$.

**3.1.2 Spectral solver direct solution compared to other theoretical published results.**

Craft solver in harmonic and parallelized version was used to perform calculations directly in frequency domain (virtual DMA). The exact same inputs have been used as those given in Masurel paper and Fig. 2 (Right) shows an example of the aleatory relaxation times assignment over the voxels used in the FFT solver. In Fig. 3 are presented the real and imaginary parts of the complex modulus for various cases of the ratio $G_g/G_r$ as published by Masurel (dotted curves) and the present results obtained with a 3D spectral solver (solid lines). The results are very close also some differences can be observed for large $G_g/G_r$ ratio which constitutes the most severe case to simulate. It is not useful here to discuss the origin of such discrepancies because they can be numerous. It can just be underlined that

i) the spectral solver is much more adapted to the multi-material case presented here than FE simulations as the spectral solver directly proceeds on the cubic (or squared) voxels. There is no error introduced here through some meshing of FE type (edge effects, location of quadrature nodes…) as the geometry of the problem is perfectly respected into the FFT scheme. This means that no errors can be made that are connected to the spatial resolution considered. In the FE approach, on the contrary, the authors had to perform a preliminary sensitivity analysis to obtain a representative multi-material which preserves accessible computation times. No such insurance is required in the FFT approach.

ii) as said before, the procedure to obtain the complex modulus is direct with the FFT solver and indirect in the FE approach, which must be coupled to an identification of a discretized weighted function.

Some errors are introduced in each of this feature specific to the FE approach and precludes any further investigation at this stage. Note that 3D simulations were probably beyond reach with FE approach whereas they can be done in very reasonable CPU times with the FFT solver.

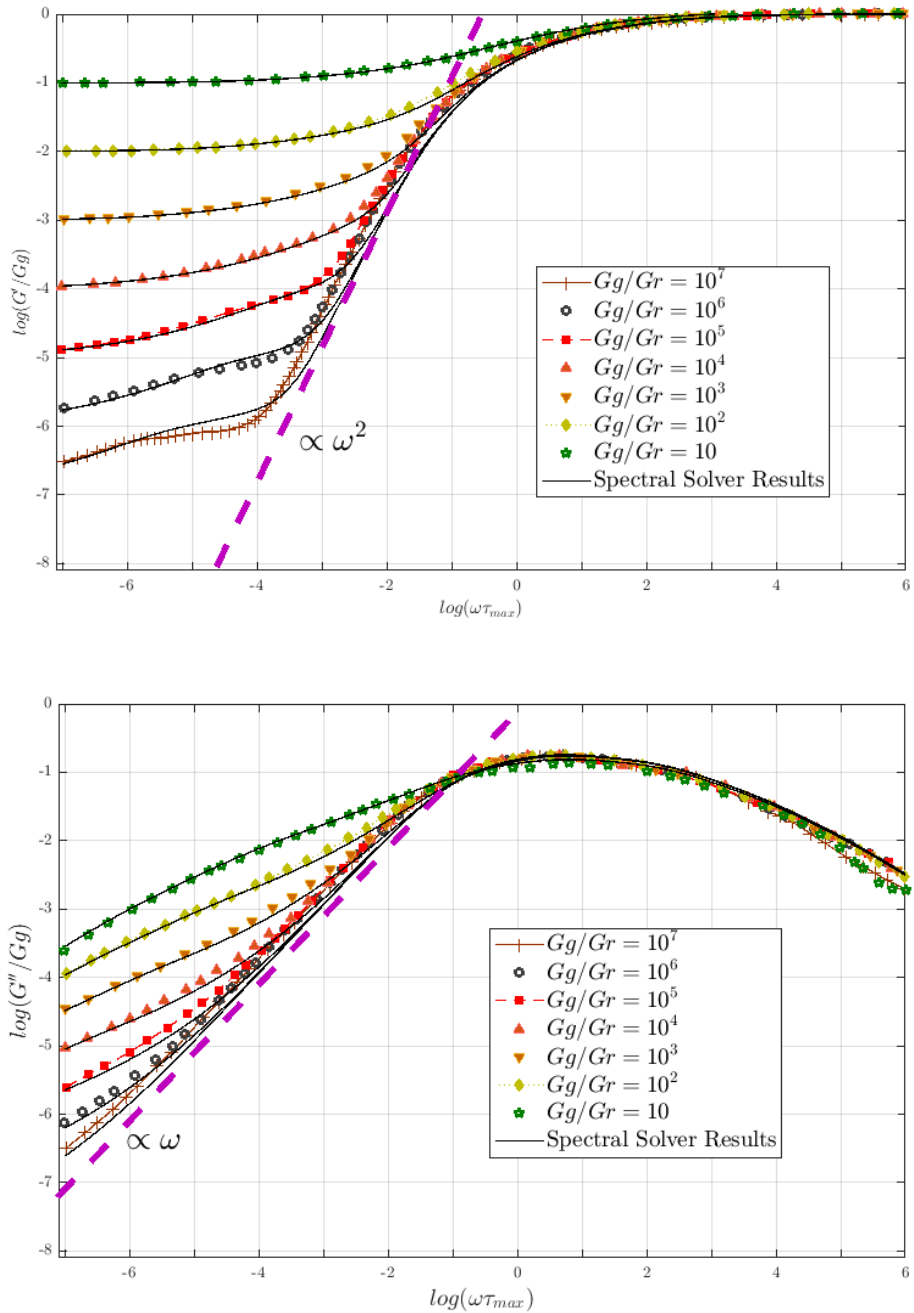

Figure 3: Storage (upper fig.) and dissipative (lower fig.) moduli as function of frequency for different contrasts $G_g/G_r$ in the glassy and relaxed moduli. Spectral solver results shown in solid lines, Datas from Masurel et al. shown in dots. Asymptotic behaviors for vanishing relaxed state shown in dotted lines.

### 3.1.3. Interest of virtual DMA

The direct results of the virtual DMA spectral solver can be very useful in such kind of applications to gain insight about physical theories of disordered structures. The multi-material proposed by Masurel and co-workers aimed at offering some alternative modeling of disordered microstructures in order to understand complex cooperative phenomena during a relaxation test (of any kind). The DMA observables virtually obtained here can now be confronted to physical models established mainly from microscopic and statistical considerations in order to see if some consistency can be achieved from both approaches. Such types of physical models are plethoric in the literature and for a pure pedagogical objective, we will here focus on two of them: the HN model (Havriliak and Negami 1995) and the JPC model established by Johari, Perez and Cavaillé (Cavaillé at al. 1989).

<u>HN model:</u> Initially proposed on empirical basis as a fitting model for DMA outputs (real and imaginary components of the material relaxation function in frequency domain) obtained on complex systems, this model further receives physical significance from probabilistic approaches to relaxation phenomenon (Jurlewicz and Veron 2002).

The HN relaxation function has the following analytical expression

$$G_{HN}^*(\omega) = \frac{1}{(1 + (i\omega/\omega_M)^\nu)^\gamma} \tag{13}$$

where $\gamma, \nu$ are non-integer coefficients.

<u>JPC model:</u> This model comes from a molecular kinetic theory for the rheology of glass, which is considered valid in its extension to amorphous polymers and which is two-sided:

- Physical mechanisms associated to uncorrelated rotational-transitional motions of molecules within a frozen bath of randomly distributed high energy sites. Assumed of a fast character, they generate an anelastic deformation which characterizes the so-called β relaxation.
- Physical mechanisms associated to cooperative, but hierarchically constrained, motions of cluster of molecules (referred to as shear microdomains in the paper for additional reasons). Assumed of slow character, they generate an irrecoverable deformation which characterizes the so-called $\alpha$ relaxation.

These mechanisms are fairly well recognized in the glass material scientific community, and have obviously been considered by Masurel et al. to propose such kind of hetero-dynamics modelling. It is then naturally considered here as capable to treat relaxation correctly regardless of the precise nature and scale of local interactions. As it offers a direct formulation of real and imaginary moduli as function of frequency, we can illustrate the interest of virtual DMA computations to confirm or validate assumptions introduced in physical approaches. The rheological equations are

$$G'_{JPC}(\omega) = \frac{J'(\omega)}{J'(\omega)^2 + J''(\omega)^2} + G_R \tag{14a}$$

$$G''_{JPC}(\omega) = \frac{J''(\omega)}{J'(\omega)^2 + J''(\omega)^2} \tag{14b}$$

With $J'(\omega)$, $J''(\omega)$ being the sum of two contributions ($\beta$ and $\alpha$ relaxations)

$$J'(\omega) = J'_\beta + \frac{1}{G_g}\left[1 + (\omega\tau_m)^{-\kappa}\cos\frac{\kappa\pi}{2} + C(\omega\tau_m)^{-\chi}\cos\frac{\chi\pi}{2}\right] \tag{15a}$$

$$J''(\omega) = J''_\beta + \frac{1}{G_g}\left[(\omega\tau_m)^{-\kappa}\sin\frac{\kappa\pi}{2} + C(\omega\tau_m)^{-\chi}\sin\frac{\chi\pi}{2}\right] \tag{15b}$$

Where $J'_\beta, J''_\beta$ are given by Eqs.(27) in the paper of Cavaille et al. 1989 along with all additional necessary information. Formally, this model reaches an amount of 10 parameters which would necessitate a very detailed sensitivity analysis study if any strong and metrological application to real data is required. Here we will perform an inverse parameter estimation without any prior sensitivity analysis considering that

i. some of them are known ($G_g$, $G_r$ that can be identified from low and high frequency asymptotes of the real and imaginary moduli but are known here as the inputs of the virtual DMA computations, $T_g$ the glass transition temperature of the polymer)
ii. the others respect realistic physical numerical values.
iii. no noise was added to virtual DMA data

Figure 4 confronts both HN and JPC models in their adjustment performance to the virtual DMA results obtained on the multidomain material and for the case $G_g/G_r = 10^3$. Adjustment of the models to the virtual DMA data is obtained through a Least-Square minimization of the residuals norm, considering either a simplex or Levenberg-Marquardt algorithm in order to check the good behavior of the parameter estimation process (sensitivity to initial values, stability of the results). Input parameters for virtual DMA simulations of the elastic unrelaxed $G_g$ and relaxed $G_R$ moduli (line (1) – Table 2) are considered as known parameters in the P.E.P. for both the HN and JPC models. As a result, all supplementary identified parameters are involved in the sole modeling approach of the relaxation(s) kinetics (lines (2) and (3)). Note that in the JPC model, the parameter $\tau_{max}$ refers to the kinetics of the β relaxation, the characteristic time associated to the $\alpha$ relaxation being explicitly absent from the parameters of the model but implicitly present and related to it through an assumption that parameter $\tau_\alpha = \tau_{max\beta}(A_\alpha G_g/\kappa)^{-(1/\kappa)}$ (see eq.(48) in Cavaille et al. 1989).

Figure 4a) plots the real and imaginary components of the complex equivalent modulus at convergence of the PEP as a function of frequency. Figure 4b) shows the corresponding Cole-

Cole representation. It is clear that the HN model is unable to describe the outputs of virtual DMA simulations. The Cole-Cole plot especially brings out this deficiency. On the contrary, the JPC model captures the heterodynamics approach much more precisely. Parameters identified for the JPC model (line (3) of Table 2) can be seen to be very close or at least in the same order of magnitude as those considered for the simulation of the generic amorphous polymer in Cavaille et al. (1989). This finally confirms that the ingredients used to simulate virtual DMA on a glassy system model are in phase with the ideas of physicists on the underlying mechanisms and demonstrate from our point of view all the benefits that can be drawn from spectral solvers used in virtual DMA mode.

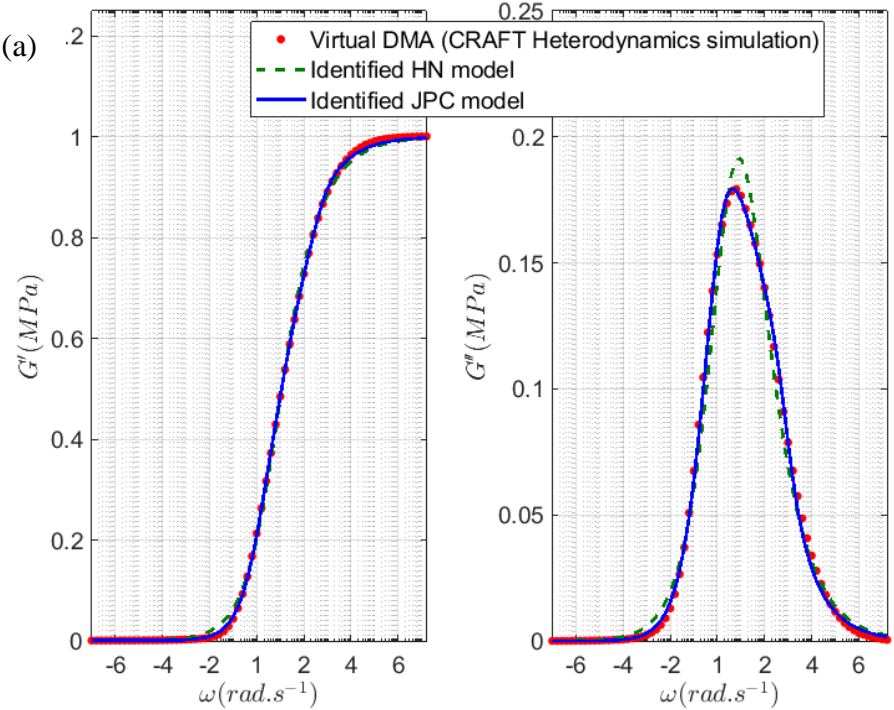

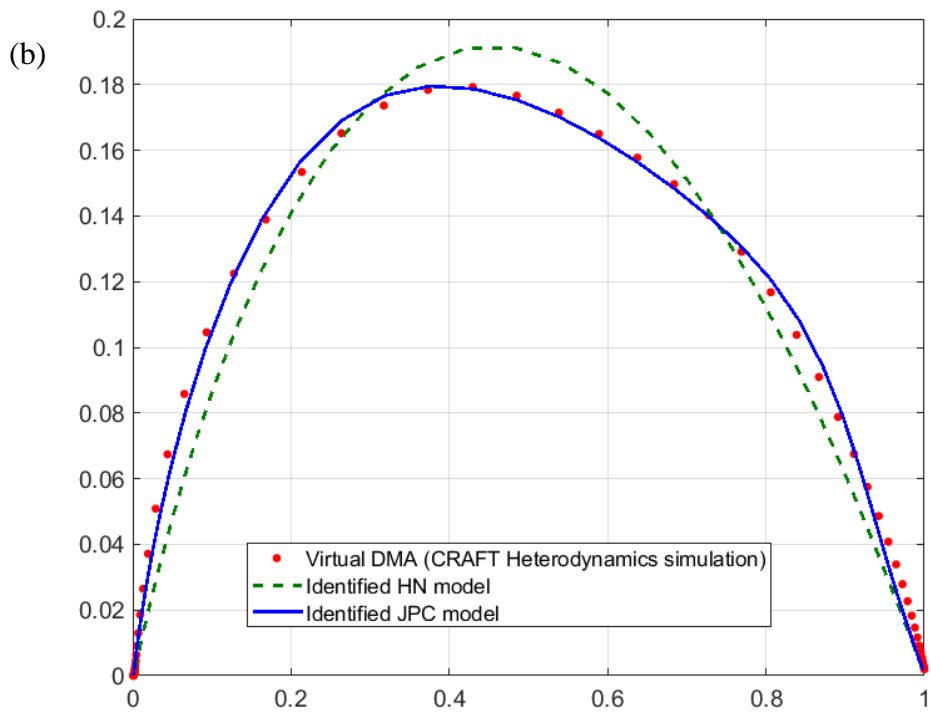

Figure 4: Virtual DMA outputs on the heterogeneous dynamics system: a) Storage and dissipative moduli – b) Corresponding Cole-Cole plot

| Virtual DMA data | | | | | | | |
|---|---|---|---|---|---|---|---|
| Input data | $G_g (GPa)$ | $G_r (GPa)$ | $\tau_{max}(s)$ | $s$ width of the normal distribution of times $\tau_i$ | | | |
| (1) | 1 | $10^{-3}$ | 1 | 2 | | | |
| HN model | | | | | | | |
| Parameters considered as known | $G_g (GPa)$ 1 | $G_r (GPa)$ $10^{-3}$ | | | | | |
| Identified Parameters | | | $\tau_{max}$ $(2\pi/\omega_{max})$ | $\nu$ | $\gamma$ | | |
| (2) | | | 0.244 | 0.532 | 0.678 | | |
| JPC model | | | | | | | |
| Parameters considered as known | $G_g (GPa)$ | $G_r (GPa)$ | $T_g (K)$ | | | | |
| | 1 | $10^{-3}$ | 355 | | | | |
| Identified parameters | $\tau_{max}(s)$ | $\kappa$ | $\chi$ | $A_\alpha^{-1}$ $(GPa^{-1})$ | $A_\beta^{-1}$ $(GPa^{-1})$ | $U$ $(kJ/mol)$ | $\Delta U$ $(kJ/mol)$ |
| (3) | 0.156 | 0.37 | 0.75 | 0.29 | 0.21 | 55 | 25 |
| (4) | 8 | 0.3 | 0.95 | 0.3 | 0.1 | 60 | 5 |
| Values considered in Cavaille et al.,1989 for a generic amorphous polymer | | | | | | | |

Table 2: Input parameters for the virtual DMA heterodynamics approach and the parameter estimated for the HN and JPC models.

**3.2: Virtual versus experimental DMA on polystyrene/glass beads composites.**

This second example concerns now a material which is a particulate composite made of glass beads (GB) considered to be elastic and embedded in a viscoelastic matrix of Polystyrene (PS). Experimental DMA results have been obtained with a temperature sweep at a given frequency. A temperature variation directly affects the relaxation time spectrum of the matrix and the overall composite behavior. Around the glass transition, microstructural configuration changes occur and probing kinetical response times makes DMA a very sensitive and adapted tool. All experimental data collected for this part were published in a series of 3 papers by Alberola and collaborators (Agbossou et al. 1993, Alberola and Bergeret 1994, Alberola and Mele 1996) and reported in terms of storage modulus and loss tangent variables. As a consequence, this example will first show how the virtual DMA code can work with tabulated pairs (real and imaginary parts) informing about the material behavior at each temperature. Secondly it will again illustrate the efficiency of a spectral code to produce either 2D or 3D DMA simulations for any type of composites from the knowledge of the individual constitutive materials. Finally, it will show that virtual DMA, as a computational tool applying on composite microstructures which mimic the topological reality in the best way, can either lead to refute previous theories or effective models or to cast doubt on the experimental data (which can advantageously lead to reconsider the metrology associated to its apparatus or the possible source of bias).

3.2.1. Data brought by experimental DMA.

The DMA experiments on PS/GB particulate composites were conducted on a Metravib equipment. The aim was to check the modeling abilities of homogenization approaches (the GSC or Generalized Self-Consistent approach). These later were based on available approaches in the elastic framework and extended to viscoelastic behavior through the correspondence principle.

In figures 1 and 7 of the 1993 paper, DMA data are given for neat PS in terms of storage modulus, $\tan\delta$ and complex Poisson coefficient respectively. These data have been digitized to provide $E^*, \nu^*$ or after conversion $K^*, \mu^*$ for implementation in CraFT solver under table format. Of course, errors have certainly been introduced in this process. The main outputs of virtual DMA reported here will show that they have no impact by themselves on the results, those latter being rather fully determined by the extrapolation made by Alberola and coll. to produce experimental Poisson coefficient values.

These data are available in the 25-185°C temperature range, 5-100Hz frequency range, for composites with $f_v = 6\%, 15\%, 21\%, 35\%, 50\%$ volume fraction of glass beads, available in two different size distributions : $d_1 = 1 - 45\mu m$ and $d_2 = 70 - 110\mu m$. Unfortunately, no precise information is known regarding the real topology of the microstructures obtained from these distributions. At large volume fractions, aggregates or clusters of particles can have been formed, with unknown packing arrangement, in possible anisotropic configurations…

(Alberola and Mele 1996). The authors also mentioned an existing porosity of about 8 vol % in the specimen. This of course precludes perfectly realistic simulations, yet easily achieved with spectral solvers. Virtual DMA simulations will then be performed assuming a given topology and as a result some discrepancies can be expected. Regarding the constituent mechanical properties, ultrasonic measurements have provided elastic constants of the polymer matrix with assumed high accuracy ($E = 3.69\ GPa, G = 1.38\ GPa, \nu = 0.33$) and at room temperature the DMA results are in accordance with them. Note that DMA technique provides only indirectly the intrinsic moduli through overall rigidity measurements, which depend on the type of mechanical configuration (tension, flexion) and on length dimensions. In the absence of any details about that point, we observe that the authors report a good consistency between DMA data and US measurements. The glass transition temperature of PS is about 100°C. Regarding GB, elastic constants are given as $E = 73\ GPa, \nu = 0.2$.

Virtual DMA isochronal scans simulations are performed directly from the knowledge of the pair $E', tan\delta$ versus temperature at 5Hz for the PS matrix (as measured with DMA by Agbossou et al. 1993, Fig. 1-a, Fig.1-b), and the knowledge of a given distribution size ($d_1$ or $d_2$) and given volume fraction of glass beads $f_v$. Note that the $log(E'')$ curve was not made available by the authors for pure PS but available for all types of composites. As a result, and because $tan\delta$ tends asymptotically to 0 at low temperatures (At 5Hz and far from the transition region, it is expected to have a pure elastic behavior), the data collected from the figures are obviously affected by artefacts which will impact the calculated imaginary moduli of the composites.

3.2.2. Conclusions drawn from Alberola and collaborators.

In the above-mentioned papers, experimental data deliver the following observations:

As the volume fraction of particles increases (from 0 to 50 % in their study):
① the magnitude of mechanical relaxation characterized by the maximum value of the $tan\delta$ peak decreases;
② a shift of the $tan\delta$ peak towards higher temperatures is reported. This effect is not obvious from the data (Fig.3-c, Agbossou et al. 1993) and could result from a small experimental bias (with regard to the temperature sweep and the thermalization of the specimen for example).
③ both glassy and relaxed moduli at low and high temperatures respectively show an increase.

At constant volume fraction,
④ the increase in average size of the particles ($d_1 \rightarrow d_2$) results in an increase in the $tan\delta$ peak maximum (dissipation), or the reinforcement effect increases with decreasing the size of the particles. We have $\left(E_{v_T}\right)_{d_1} > \left(E_{v_T}\right)_{d_2}$ and $(tan\delta)_{d_1} < (tan\delta)_{d_2}$. Size effects depend obviously on the nature of the matrix/fillers couple and the review paper of Fu et al (2008) reports for example that the modulus of a composite with high volume fraction of GB in an epoxy resin is nearly insensitive to size effects.

A GSC mean field theoretical approach (Kerner's and Christensen and Lo's models for the bulk and shear moduli respectively) was used in complex variables to explain these data. Only the observed fact ③ was reproduced by the model, based on a constant and real valued Poisson ratio (whatever its value between the low temperature $\nu = 0.33$ and high temperature $\nu \to 0.5$). Especially expected behaviors ① and ② escape to the models: drop in maximum and shift in temperature of $tan\delta$ with an increase in volume fraction of GB (See Fig 5 and 6 in Agbossou et al. 1993). This led the authors to invoke a complex-valued Poisson coefficient and a strategy to determine it on the considered temperature range. This point will deserve a large discussion in the next section.

3.2.3: Virtual DMA results and implications regarding a comprehensive analysis of the role played by the Poisson ratio of the matrix.

DMA simulations of the particulate composite are based on a synthetic material created with the Random Sequential Algorithm. For large volume fractions (up to 50%), a metropolis algorithm is combined to it to reduce computation times. Note that without additional reported information, the two size distributions of the particles are considered as gaussian. The chosen $R_{max}/R_{min}$ ratio and volume fraction $V_f$ allow distributing a number of spheres $N$ of mean radius $r_{mean}$ in a square box of unit dimensions.

**Influence of the volume fraction of the particles**

Because this point was the major fact discussed in Agbossou et al. (1993), we first present some results obtained with the virtual DMA solver along with the available experimental data. Virtual DMA simulations have been performed for 3 different values of the matrix Poisson coefficient $\nu_1 = 0.33$, $\nu_2 = 0.49$, $\nu_3 = 0.4999$ assumed constant (independent of temperature). Note that the values 0.49 and 0.4999 are selected to represent the incompressible behavior expected in rubbery state (at high temperature) but without any possible precise experimental confirmation. Figures 5-a and 5-b compare experimental and simulated data for the storage modulus and the loss angle tangent in the case of a size distribution $d_1$ The following observations can be made (same as for distribution $d_2$):

- For a volume fraction of 6% of particles, virtual DMA solutions are very close to the experimental data for the pure PS matrix. The latter data are used as input for the matrix properties so that this result is expected. Any mean-field approach would lead to the same results, as a matter of fact for small concentrations. Note that experimental data (full circles) nevertheless departs substantially from the matrix data. This naturally results from a mismatch between the intrinsic ideal character of theoretical models (whatever they are) and the experimental conditions actually followed. On one hand, the deficiencies in the model could be relative to a lack of knowledge on the specimen (porosity, agregates, interphases, statistical distribution of sizes unknown…). On the other hand, one could incriminate biases in the experimentation. Regarding DMA one has to keep in mind that intrinsic measurements, like the real and imaginary moduli provided here, are obtained through an overall stiffness measurement. They do already depend from an identification model and its assumptions (stability in size, loading conditions through grips…). Because this discrepancy arises only in the transition range between 110°C-130°C, a bias due to the temperature regime could also be suspected, especially in the temperature sweep mode.

- For a volume fraction of 50% of particles, virtual DMA solutions produce a $tan\delta$ curve very close to the experimental ones with a strong peak drop at the transition temperature (compare to Figs 5 and 6 of the cited paper). The computations show also a very high sensitivity to the value of the Poisson ratio: considering $\nu_2$ or $\nu_3$ values for the incompressible state has a drastic impact on the solution. The same sensitivity was shown previously when computing bounds of the homogenized bulk modulus (Fig. 4 in Gibianski and Lakes, 1993). These two results clearly escape to the mean-field approach ('GSC model' curve reported in Fig. 5b, quasi-insensitive to the Poisson ratio). In virtual DMA solutions with constant real Poisson ratio, experimental data are well described by the value $\nu_2 = 0.49$ for both the storage modulus and loss angle. It is worthwhile to note that this sensitivity logically begins only as the material enters into the vitreous transition i.e. above temperatures of 110°C, when the mechanical properties of the PS matrix drastically decreases towards the rubber state. Thus, the value of 0.49 is very sounded. The temperature dependence of the Poisson ratio is taken into account in the next subsection and confirms that point.

- The increase in volume fraction of glass beads makes both glassy and rubber moduli greater at low and high temperature respectively. This fact was the only one observed similarly with mean-field GSC approaches (('GSC model' curve reported in Fig. 5a). But DMA simulations produce a slightly higher reinforcement at low temperature compared to the experiment. The porosity of 8% considered by the authors as a result of the composite elaboration process can explain this fact. Note that according to theoretical simulations of Remillat (Remillat 2007, Fig.2), our ratio of elastic moduli particle/matrix is 20 and we would expect at this volume fraction of particles a 2.5 factor gain in stiffness. Virtual DMA gives an homogeneized elasticity of about 9.8 GPa at 60°C for a 50% volume fraction which falls exactly as $2.5 \times 3.7 GPa$. This again could confirm a pretty good consistency of the virtual DMA predictions and consequently that the discrepancies with experimental data may primarily be due to a specimen microstructure mismatch between idealized and real composites. Regarding the behavior of the storage modulus at high temperature, it can be clearly seen in Fig.5a that the GSC model unerestimates its value when compared to virtual DMA.

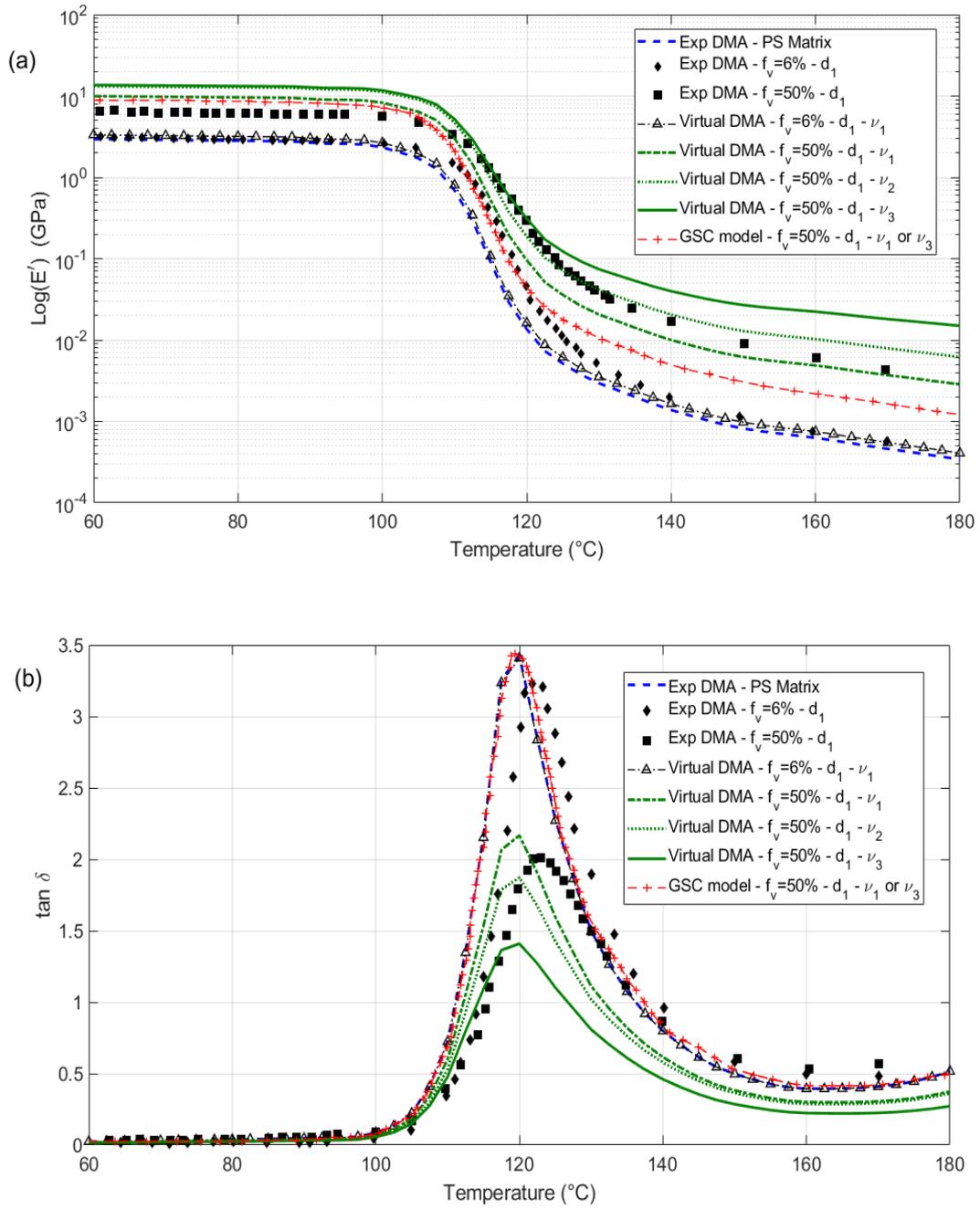

Figure 5: Storage modulus (a) and Loss tangent (b) versus temperature – Distribution $d_1$.

Virtual DMA thus directly confirms observations ①,③ without any requirement of a Poisson coefficient being a complex quantity, the central argument of the paper of Agbossou et al. (1993), to correct the model-experiment inconsistency at high volume fraction of particles. But still, their 2-phase model (Christensen and Lo 1979) clearly overestimates the maximum of $tan\delta$ peak (Fig. 8-c, ibid and Fig.5b – Curve GSC model). Later (Alberola and Mele 1996), observation ① was met theoretically using a four-phase model (Hervé and Zaoui 1993), invoking possible aggregation of particles at high volume fractions (50%) but considering a Poisson coefficient as a real quantity. Again, it is clear that virtual DMA strongly questioned this assumption as it is not required here in our simulations. The better agreement obtained by the authors between experimental data and their model probably just results from an increase in the degrees of freedom of the model, going from the two-phase to four-phase self-consistent approach. Note also that the complex-valued Poisson coefficient extrapolated by Agbossou et al. (1993) for the whole considered temperature range, when converted into bulk and shear moduli (for introduction in the behavior's law of Craft solver), produces a negative real bulk modulus which is not physical and probably escape the authors attention.

Regarding observation ② (slight phase shift towards higher temperatures with the increase in $f_v$), it is not confirmed by the virtual simulations. The peak location of the composite is totally determined by the $tan\delta$ peak of the matrix (input data). This is an obvious consequence of the temperature determined physical transition of the matrix only and to the known fact that a high damping of the compliant phase gives rise to high damping in the composite in spite of the fact that the moduli of this phase are much smaller in absolute value than the moduli of the stiff elastic phase (Gibianski and Lakes 1997). The loss factor curve of the mixture is ruled by that of the matrix, regardless of the volume fraction, if the inclusion is either very stiff or very soft. This favors our hypothesis that this effect is not so tangible and therefore, that experimental results might have a bias.

**Influence of the size distribution of the particles**

Simulations shown in Fig. 6 were performed for the two size distributions of the particles $d_1$ (open symbols) and $d_2$ (full symbols). Regarding experimental observation ④, virtual DMA simulations contradict this point. In the case of constant Poisson ratio for the matrix equal to 0.49, we observe from the simulations that an increase in average size of the particles at the same constant volume fraction ($f_v = 50\%$) increases the reinforcement: $(E_{vT})_{d_1} < (E_{vT})_{d_2}$ and $(tan\delta)_{d_1} > (tan\delta)_{d_2}$. This is the exact contrary of what is reported experimentally. Note that this question could not have been adressed by Alberola and collaborators within the Self-Consistent models they used (taking into account a volume fraction parameter but no parameter relative to any particle size). For the two specimen at 50% volume fraction and distributions $d_1$ or $d_2$, the conclusion is then that the model can not explain the data when an equal matrix Poisson ratio behavior is considered.

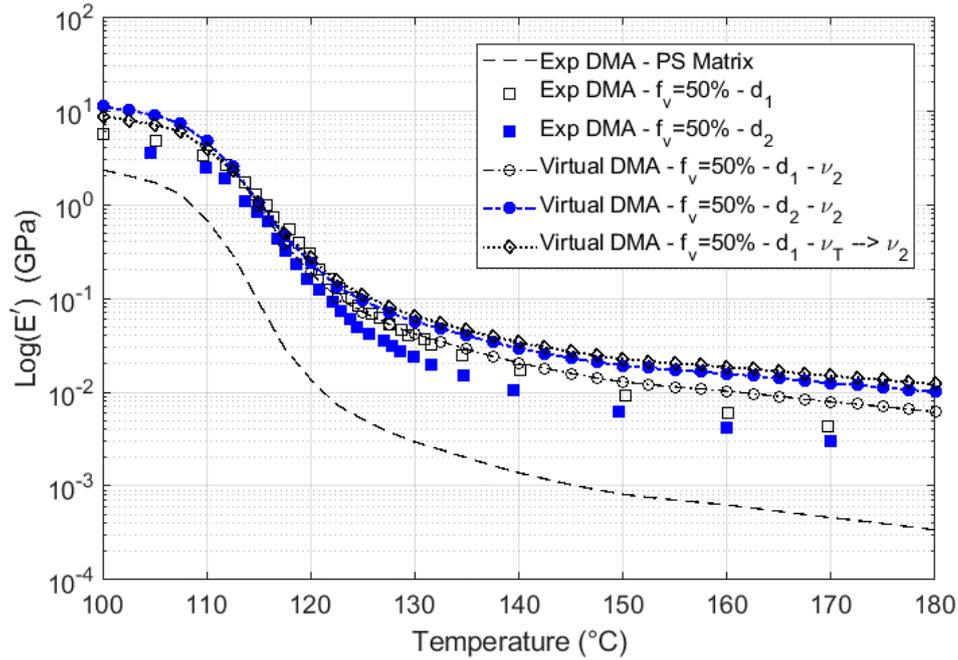

Figure 6: Storage Modulus (GPa) versus temperature –Distributions $d_1$ and $d_2$ for the same volume fraction of 50%.

With some reasons given previously, an experimental bias could be invoked to explain this conflict. But because it is clear from Fig.6 that the sensitivity to the Poisson ratio is of crucial concern, the Poisson ratio of the matrix is now considered to vary with temperature (noted $\nu_T$ in that case) according to a S-shape function, starting at $\nu_T = \nu_1$ at room temperature, and with an asymptotic value $\nu_T$ at high temperature ($180°C$) that is supposed to tend towards ½ because the matrix tends to an incompressible behavior. This fact was demonstrated experimentally long times ago in the works of Kono (Kono 1960) and Waterman (Waterman 1963) with $\nu_T$ of PolyStyrene considered to vary from 0.35→0.5 or 0.3→0.45 respectively for the two authors. If we apply the same procedure as in Agbossou et al. to reconstruct $\nu_T$ from experimental DMA data, with the constraint $\nu_T \to \nu_2 = 0.49$ and $\nu_T$ always considered as a real-valued quantity, it can be seen in the plots of Fig. 6 that the extreme sensitivity to the Poisson ratio could explain such conflict: for a same volume fraction of GB, a Poisson ratio considered as varying with temperature and $\nu_T \to \nu_2 = 0.49$ for distribution $d_1$ leads to an overall curve of rigidity now surpassing the curve obtained for distribution $d_2$ with a Poisson ratio considered as constant and equal to $\nu_2 = 0.49$. DMA experimental data have been obtained up to 190°C, a temperature still very far from the fusion temperature where a drastic drop in modulus occurs. Hence the rubbery state can be more or less established. Additionnaly, DMA experiments conducted under temperature sweep condition impose a given temperature rate. Specimen with different size distribution for the glass beads can easily produce slightly different thermal conductivity, a parameter which can change substantially the thermal regime in sweeping mode and hence change the behavior with respect to the incompressibility limit. The presence of voids in the highly concentrate case (already discussed) can also affect this parameter. Anyway, if the reasons explaining this possible conflict cannot be firmly established,

<span style="color:red">on the other hand,</span> this issue shows how virtual DMA simulation can contribute to open avenues for reflection. In the present case, it suggests strong investigations relative to the precise measurement of Poisson ratio of the matrix with temperature, especially when it tends to a rubbery state.

Finally, as was shown for example 1, outputs of virtual DMA in terms of mechanical fields can be valuable to get deeper analysis of phase transition. We select in Fig. 7-a the volumic strain fields of the matrix, calculated for the same parameters as for Fig.6. Two maps are shown, obtained for a temperature below (60°C) and above (160°C) the glass transition temperature $T_g$ and for an input (average) stress of 1. These fields are shown here normalized with respect to the longitudinal modulus as it changes drastically when crossing the transition zone. When compared to the map at 60°C, it can be seen that the volumic strain obtained above $T_g$ tends to a more homogeneous zero-value everywhere in the matrix, except in the very limited area confined between two beads. Note that in this simulation, the Poisson ratio tends 'only' to 0.49. Figure 7-b shows the overal volumic strain averaged over a selection of 10 different configuration along with the volumic strain in both the GB and PS matrix phases. The transition regime marks the decrease of both volumic strains. In the GB, it tends to zero within the numerical precision when raising the temperature. In the matrix, it decreases strongly of nearly one order of magnitude at the transition. It slightly increases beyond probably because the matrix sustains the all imposed mechanical excitation and as shown in the strain maps, strong volume expansion may exist locally because the Poisson ratio is not exactly equal to 0.5.

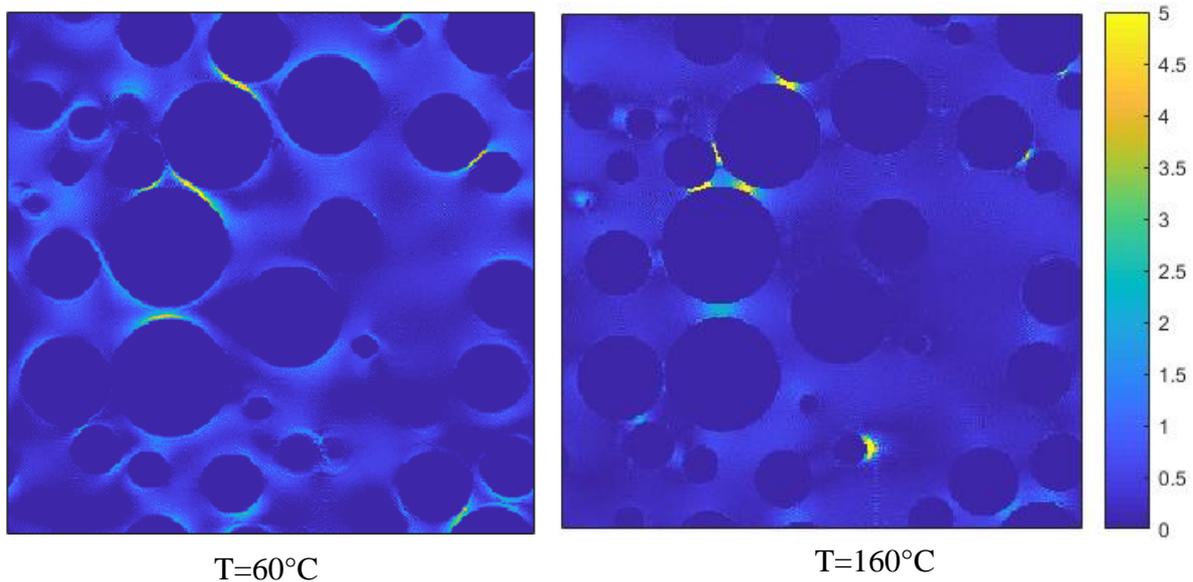

T=60°C  T=160°C

Figure 7 :
a- Volumic strain maps at T=60°C and T=160°C (same input data as for Fig.6).
b- Volumic strain evolution with temperature (overall, in the PS matrix, in the GB particles) – Averaged over 10 configurations.

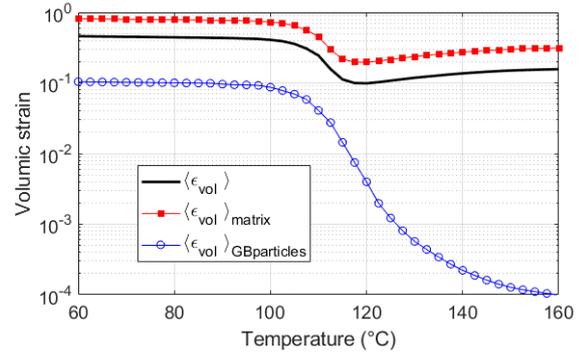

## 4. Conclusion

The study presented here was intentionally based upon data provided in previously published papers, either theoretical or purely experimental, to show the potentialities of virtual DMA. As a simulation tool based on an FFT solver, it turns out to be a very competitive alternative path to classical FEM in terms of computation times. For experimentalists, it offers a direct access to the frequency-dependent observables of this widely used technique: storage and dissipative moduli, loss tangent. For a given heterogeneous RVE, it provides at the same time the full-field strain or stress maps, in either 2D or 3D, which can be valued to push forward our understanding of its mechanical behavior. Due to a favorable precision/computation times ratio, it is expected that in a near future, such codes will be able to perform inverse parameter identification on the collected experimental data. Hence it will help to question the validity of homogenized models or the experimental technique itself with regards to the way data are produced. But one of the most interesting perspective that virtual DMA offers is maybe the possibility to check the consistency of an idealized microstructure with respect to viscoelasticity. Based on MEB, X-ray tomography, AFM or any other imaging technique of the microstructural organization of some material, the complementary observation of the macroscopic behavior of this microstructure in a DMA experiment, both real and virtual, will allow fixing this microstructure idealization as consistent either in terms of morphological parameters (volume fraction of constituents, aspect ratio of the phase…) and mechanical parameters (moduli, Poisson ratio, relaxation time spectra…). Considering such a way to firmly assess the microstructure model, routes will then be open to consider the variety of expected behaviors at a larger deformation range (plasticity, hardening, damage…).

**Acknowledgments:** The authors wish to thank Hervé Moulinec (Laboratoire de Mécanique et Acoustique - Marseille - UMR CNRS 7031) for a fruitful introduction to CraFT and guidance in using it.